\documentclass[A4,11pt,epsf,oneside]{article}
\usepackage{amssymb, amscd, amsmath,amsthm,amsfonts,graphicx}
\usepackage{epsfig}
\usepackage{xcolor}
\usepackage{capt-of}
\usepackage{float}
\restylefloat{table}

\begin{document}
\renewcommand{\thefootnote} {\fnsymbol{footnote}}
\setcounter{page}{1}

\title{\textbf{Plane Symmetric, Cylindrically Symmetric and Spherically Symmetric Vacuum Solutions of Einstein Field Equations}}
\author{\vspace{0.5cm} {\bf{Farhad Ali}}
\\
School of Natural Sciences, \\
National University of Sciences and Technology, \\
H-12 Islamabad, Pakistan. \\Email: farhadmardan@gmail.com }
\maketitle
\begin{abstract} In this paper we present Plane symmetric, Cylindrically Symmetric and Spherically Symmetric Black hole or Vacuum solutions of Einstein Field Equations(EFEs). Some of these solutions are new which we have not seen in the literature. This calculation will help us in understanding the gravitational wave and gravitational wave spacetimes.\end{abstract}

\section{Introduction}
General plane symmetric \cite{f}, cylindrically symmetric and spherically symmetric static spacetimes are consider for calculating the vacuum solutions of EFEs \cite{Q}. The calculation is straight forward, we calculate the Ricci tensors of the generally plane symmetric, cylindrically symmetric and spherically symmetric static spacetimes and put these Ricci curvature tensors equal to zero. Obtaining a system of three non-linear partial differential equations in plane symmetric case, four non-linear partial differential equations in cylindrically symmetric and three non-linear partial differential equations in spherically symmetric static spacetimes. The solutions of these system give us the required vacuum solutions of EFEs in each case. In all these calculation we are searching all those spacetimes which are the vacuum solutions or black hole solutions of EFEs \cite{g}. In all our calculation we have seen only one singularity which is the essential singularity and occur at $r=0$. These spacetimes will help in understanding of the gravitational wave spacetime, black hole \cite{R} and asymptotic behavior of black hole \cite{p}.
\section{Plane symmetric Static Spacetimes and Vacuum solutions of EFEs}Consider the following general plane symmetric static spacetime \begin{equation}ds^2=e^{\nu(x)}dt^2-dx^2-e^{\mu(x)}(dy^2+dz^2).\label{1}\end{equation}
We find the Ricci curvature tensors for this spacetimes and put them equal to zero. We get a system of three non-linear partial differential equations in two unknown functions $\nu(x)$ and $\mu(x)$,\begin{equation}\begin{split}&R_{00}=2\nu_{xx}(x)+2\nu_x(x)\mu_x(x)+\nu_x^2(x)=0,\\&R_{11}=2\nu_{xx}(x)+4\mu_{xx}(x)+\nu_x^2(x)+2\mu_{x}^2(x)=0,\\&
 R_{22}=2\mu_{xx}(x)+2\mu_x^2(x)+\nu_x(x)\mu_x(x)=0.\end{split}\label{2}\end{equation} The solution of
this system is\begin{equation}\nu(x)=-\frac{2}{3}\ln\big(\frac{x}{\alpha}\big),\quad\mu(x)=\frac{4}{3}\ln\big(\frac{x}{\alpha}\big).\label{8}\end{equation}
Which is the famous Taub Spacetime. This is a static gravitational wave spacetime and we check that it does not admit time conformal perturbation to form it an actual non-static gravitational wave spacetime \cite{z}. We can see that there is only one singularity at $r=0$. For time conformal perturbation either one of the exponent of $\frac{x}{\alpha}$ is $2$ or  $\nu(x)=\mu(x)=a\ln\big(\frac{x}{\alpha}\big)$.  Which mean that the metric for plane symmetric vacuum solutions of EFEs must be static and independent of time as was proved by Taub in his paper \cite{T}.

\section{Cylindrically Symmetric Static Vacuum Solution of EFEs}
The general metric of cylindrically symmetric static spacetimes is \cite{Q}
\begin{equation}ds^2=e^{\nu(r)}dt^2-dr^2-e^{\mu(r)}d\theta^2-e^{\lambda(r)}dz^2.\label{3}\end{equation}
We calculated the Ricci tensors of the spacetimes given in equation (\ref{3}) and put them equal to zero we get the following system of four non-linear partial differential equations in unknown functions  $\nu(r)$, $\mu(r)$ and $\lambda(r)$.
\begin{equation}\begin{split}&R_{00}={\nu}_r^2(r)+2\nu_{rr}(r)+{\nu}_r(r){\mu}_r(r)+{\nu}_r(r){\lambda}_r(r)=0,\\&
R_{11}={\nu}_r^2(r)+2\nu_{rr}(r)+{\mu}_r^2(r)+2\mu_{rr}(r)+{\lambda}_r^2(r)+2\lambda_{rr}(r)=0,\\&
R_{22}={\nu}_r(r){\mu}_r(r)+{\mu}_r^2(r)+2\mu_{rr}(r)+{\mu}_r(r){\lambda}_r(r)=0,\\&
R_{33}={\nu}_r(r){\lambda}_r(r)+{\lambda}_r^2(r)+2\lambda_{rr}(r)+{\mu}_r(r){\lambda}_r(r)=0.
\end{split}\label{4}\end{equation}
The solution set to the system (\ref{4}) is
 \begin{equation}\begin{split}&(i):\quad\nu(r)=2\ln\big(\frac{r}{\alpha}\big),\quad\mu(r)=c_1,\quad\lambda(r)=c_2
 ,\\&(ii):\quad\nu(r)=c_1,\quad\mu(r)=2\ln\big(\frac{r}{\alpha}\big),\quad\lambda(r)=c_2
 ,\\&(iii):\quad\nu(r)=c_1,\quad\mu(r)=c_2,\quad\lambda(r)=2\ln\big(\frac{r}{\alpha}\big),\\&
 (iv):\quad\nu(r)=\frac{1+2a+\sqrt{4a^2-4a-3}}{2a}\ln\big(\frac{r}{\alpha}\big),\quad\mu(r)=-\frac{1}{a}\ln\big(\frac{r}{\alpha}\big),\quad
\lambda(r)=\frac{1+2a-\sqrt{4a^2-4a-3}}{2a}\ln\big(\frac{r}{\alpha}\big),\\&
(v):\quad\nu(r)=\frac{1+2a-\sqrt{4a^2-4a-3}}{2a}\ln\big(\frac{r}{\alpha}\big),\quad\mu(r)=-\frac{1}{a}\ln\big(\frac{r}{\alpha}\big),\quad
\lambda(r)=\frac{1+2a+\sqrt{4a^2-4a-3}}{2a}\ln\big(\frac{r}{\alpha}\big),\\&
(vi):\quad\nu(r)=-\frac{1}{a}\ln\big(\frac{r}{\alpha}\big),\quad\mu(r)=\frac{1+2a+\sqrt{4a^2-4a-3}}{2a}\ln\big(\frac{r}{\alpha}\big),\quad
\lambda(r)=\frac{1+2a-\sqrt{4a^2-4a-3}}{2a}\ln\big(\frac{r}{\alpha}\big),\\&
(vii):\quad\nu(r)=-\frac{1}{a}\ln\big(\frac{r}{\alpha}\big),\quad\mu(r)=\frac{1+2a-\sqrt{4a^2-4a-3}}{2a}\ln\big(\frac{r}{\alpha}\big),\quad
\lambda(r)=\frac{1+2a+\sqrt{4a^2-4a-3}}{2a}\ln\big(\frac{r}{\alpha}\big),\\&
(viii):\quad\nu(r)=\frac{1+2a+\sqrt{4a^2-4a-3}}{2a}\ln\big(\frac{r}{\alpha}\big),\quad\mu(r)=\frac{1+2a-\sqrt{4a^2-4a-3}}{2a}\ln\big(\frac{r}{\alpha}\big),\quad
\lambda(r)=-\frac{1}{a}\ln\big(\frac{r}{\alpha}\big),\\&
(ix):\quad\nu(r)=\frac{1+2a-\sqrt{4a^2-4a-3}}{2a}\ln\big(\frac{r}{\alpha}\big),\quad\mu(r)=\frac{1+2a+\sqrt{4a^2-4a-3}}{2a}\ln\big(\frac{r}{\alpha}\big)
,\quad
\lambda(r)=-\frac{1}{a}\ln\big(\frac{r}{\alpha}\big).\label{5}\end{split}\end{equation}The solutions (i), (ii) and (iii) are the cone solutions that is they are the flat spacetimes where all the Reimann curvature tensors are vanishing (Minkowski spacetimes) \cite{10}. The remaining solutions from (iv) to (ix) are the solutions of EFEs which have some important features. We are going to discuss some of the important points about the spacetimes. The metric for case (iv) in solution set (\ref{5}) takes the form
\begin{equation}ds^2=\big(\frac{r}{\alpha}\big)^{\frac{1+2a+\sqrt{4a^2-4a-3}}{2a}}dt^2-dr^2-\big(\frac{r}{\alpha}\big)^{-\frac{1}{a}}d\theta^2-
\big(\frac{r}{\alpha}\big)^{\frac{1+2a-\sqrt{4a^2-4a-3}}{2a}}dz^2.\label{6}\end{equation}Here the exponents of $\frac{r}{\alpha}$ are the following functions
\begin{equation}\begin{split}&f_1(a)=\frac{1+2a+\sqrt{4a^2-4a-3}}{2a},\\&f_2(a)=-\frac{1}{a},\\&f_3(a)=\frac{1+2a-\sqrt{4a^2-4a-3}}{2a}\label{w}\end{split}\end{equation}
The importance of these exponents are describe in the following lines. The functions $f_1(a)$, $f_2(a)$ are defined on the domain $(-\infty,-\frac{1}{2})U(\frac{3}{2}, \infty)$ and the function $f_3(a)$ are define for all real $a$ except at $a=0$. For $a=-\frac{1}{2}$ the solutions $(iv)-(ix)$ are the cone solutions as given in $(i)$, $(ii)$ and $(iii),$ while for $a=\frac{3}{2}$ we have the following solutions
\begin{equation}\begin{split}&(k):\quad\nu(r)=\frac{4}{3}\ln\big(\frac{r}{\alpha}\big),\quad\mu(r)=-\frac{2}{3}\ln\big(\frac{r}{\alpha}\big)
,\quad\lambda(r)=\frac{4}{3}\ln\big(\frac{r}{\alpha}\big),\\&(l):\quad\nu(r)=\frac{4}{3}\ln\big(\frac{r}{\alpha}\big),
\quad\mu(r)=\frac{4}{3}\ln\big(\frac{r}{\alpha}\big),\quad\lambda(r)=-\frac{2}{3}\ln\big(\frac{r}{\alpha}\big),\\&
(m):\quad\nu(r)=-\frac{2}{3}\ln\big(\frac{r}{\alpha}\big),\quad\mu(r)=\frac{4}{3}\ln\big(\frac{r}{\alpha}\big)
,\quad\lambda(r)=\frac{4}{3}\ln\big(\frac{r}{\alpha}\big).\label{7}\end{split}\end{equation} The coefficients given in $(m)$ are again the same as given in equation (\ref{8}), that is, it is the Taub spacetime and the coefficients given in $(k)$ and $(l)$ are new which we have not seen in the literature. It is interesting to note that for $a=-\frac{3}{4}$ we get the solution set (\ref{7}) again in different order that is for this value of $a$ we have $f_1(a)=-\frac{2}{3}$., $f_2(a)=\frac{4}{3}$ and $f_3(a)=\frac{4}{3}$.

 \section{Asymptotic behavior}For asymptotic behavior of these spacetimes we need to draw graphs of function given in (\ref{w}) and check the limits of these functions. The limiting values of $f_1(a)$, $f_2(a)$ and $f_3(a)$ are as follows, when $a\rightarrow\infty$ then $f_1(a)\rightarrow 2$, $f_2(a)\rightarrow 0$ and $f_3(a)\rightarrow 0$ while when $a\rightarrow-\infty$ then $f_1(a)\rightarrow 0$, $f_2(a)\rightarrow 0$ and $f_3(a)\rightarrow 2$ so in all the limiting cases we have the cone solutions of EFEs which are given in $(i)$, $(ii)$ and $(iii)$. Otherwise we will have the curvature in the spacetimes and will get some gravitating source in the spacetimes other than the energy momentum tensor $T_{\mu\nu}$ \cite{7}. It is evident from the graphs of the functions $f_1(a)$, $f_2(a)$  and $f_3(a)$ given in the following figures, that there is always a singularity at $r=0$ in all cases, which is the essential singularity, and there is no other singularity in these cases \cite{l}.

\begin{figure}[H]
  \centering
  \includegraphics[width=8cm]{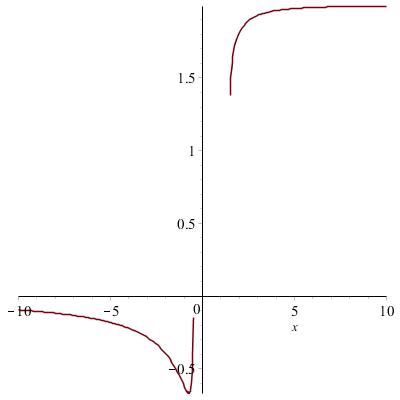}\\
  \caption{Graph of $f_1(a)$}
\end{figure}

\begin{figure}[H]
\centering
  \includegraphics[width=8cm]{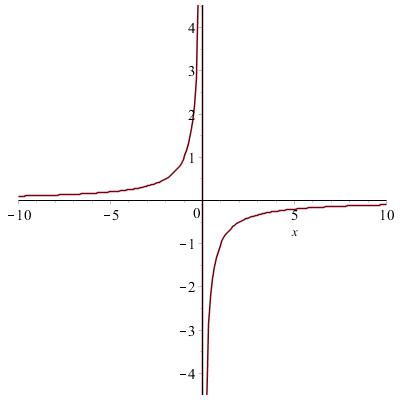}\\
  \caption{Graph of $f_2(a)$}
\end{figure}

\begin{figure}[H]
\centering
  \includegraphics[width=8cm]{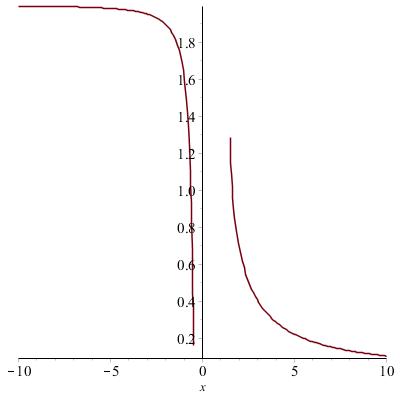}
  \caption{Graph of $f_3(a)$}
\end{figure}
Cylindrically symmetric vacuum solution of EFEs do not admit time conformal perturbation. Therefore the line element of cylindrically symmetric vacuum solution of EFEs must be in dependent of time. For the spacetime (\ref{6}) we have three possibilities to admit time conformal perturbation\\ (1):\quad $f_1(a)=f_2(a)=f_3(a)$,\\(2):\quad $f_1(a)=2$ and $f_2(a)=f_3(a)$,\\(3):\quad$f_2(a)=f_3(a)=2$.\\But it is clear from the graphs of these functions that non of these three conditions satisfied, which confirm that the line element of cylindrically symmetric vacuum solutions of EFEs are independent of time.
\section{Spherically symmetric Vacuum Solution of EFEs}
The general metric for spherically symmetric static spacetime is \cite{5}
\begin{equation}ds^2={\nu(r)}dt^2-{\mu(r)}dr^2-r^2(d\theta^2+\sin^2\theta d\phi^2)\label{9}\end{equation}For $nu(r)=\mu(r)\neq0$, obtaining the general Ricci curvature tensors for this spacetime and them equal to zero we have
\begin{equation}\begin{split}&R_{00}=-\mu(r)\nu_{rr}(r)\nu(r)r-\mu(r)\nu_r(r)^2r-\nu(r)\nu_{r}(r)\mu_r(r)r-4\mu_r(r)\nu(r)^2=0,\\&
R_{11}=2\mu(r)\nu_{rr}(r)\nu(r)r-\mu(r)\nu_r(r)^2r-\nu(r)\nu_{r}(r)\mu_r(r)r+4\mu_r(r)\nu_r(r)=0,\\&
R_{22}=-2\nu_r(r)\mu(r)r\theta^2+2\mu_r(r)\nu(r)r\theta^2-4\mu(r)\nu(r)\theta^2+\mu(r)^2\nu(r)=0.\label{10}\end{split}\end{equation}
the solution of this system is \begin{align}\nu(r)=(1-\frac{m}{r}),\quad\mu(r)=\frac{1}{1-\frac{m}{r}}\label{11}\end{align} where $m$ is an arbitrary constant. These values of $\nu(r)$ and $\mu(r)$ are define the famous Schwarzschild spacetime. For spherically symmetric static spacetime we get exactly one vacuum solution which is the Schwarzschild spacetime.
\section{Conclusion}
Here we presented plane symmetric static, cylindrically symmetric static and spherically symmetric static vacuum solutions of EFEs. In all the three cases we consider the general form the corresponding spacetimes find the general form of Ricci curvature tensors in each case and put them equal to zero we obtained system of determining partial differential equations. The solution of the determining partial differential equations provide us the require vacuum solutions of EFEs in each case. In section 2 plane symmetric static vacuum solutions are discussed. There is only one plane symmetric static vacuum solution of Einstein field equations. In section 3 we find all the cylindrically symmetric static vacuum solutions. Most of the spacetimes obtained in section 3 are new in the literature. Section 4 consist on the calculation of spherically symmetric static vacuum solution. Here we obtained only one solution which is the famous Schwarzschild solution of EFEs.

\end{document}